# Ten Puzzles in Particle Theory and Quarks as Only Possible Magic Particles with Some Symmetries

Yi-Fang Chang

Department of Physics, Yunnan University, Kunming, 650091, China

(e-mail: yifangchang1030@hotmail.com)

**Abstract:** We propose ten puzzles in particle theory: quark, asymmetries of quantum numbers in the Standard Model, point particle and interaction-distance, mystic Higgs mechanism, possible violation of the Pauli Exclusion Principle and some basic principles, neutrino and its oscillation, uncertainty principle and superstring, the superposition principle and entangled state, the wave property and duality, dark matter and dark energy. In particular, it is discussed that quarks possess many questions, and are possibly only some symmetrical magic particles.

In particle physics there are many very succeed theories, some related conclusions appear simultaneously, for example, quarks and their masses, color and confinement, the Standard Model (SM), Higgs mechanism, neutrinos and their oscillation, superstring, entangled state, monopole, various anomalous predictions and so on. But, these conclusions have some puzzles, therefore new physics and beyond SM, etc. [1-38], are proposed. They are related to electroweak interactions [1,2] and symmetry breaking, CP violation [3-5], supersymmetry, possible fourth generation of quarks and leptons [6,7], new particles and interactions (e.g., non-minimal Higgs sectors, new gauge bosons and/or exotic fermions) [8], and so on. Hurth discussed present status of inclusive rare B decays in the search for new physics [9].

The Large Hadron Collider (LHC) promises to discover new physics beyond SM. For new physics, recently CDF Collaboration searched it at CDF [10,11], and revealed no indication of physics beyond SM [10]. Faller, et al., discussed precision physics with $B_s^0 \to J/\psi$ decay products play a key role for the search of new physics at the LHC [12]. Shelton discussed that the polarized top quarks probes the chiral structure of new physics at the LHC [13]. Chang, et al., explored nonsupersymmetric new physics by the precision polarized Moller scattering experiments [14]. Hou, et al., considered the fourth generation leptons and muon g-2, and the flavor changing neutral couplings arising from various new physics models [15]. Rindani, et al., examined the contribution of general three-point interactions arising from new physics to the Higgs production process, and obtained the sensitivity of the various observables in constraining the new-physics interactions at a linear collider operating at a center-of-mass energy of 500 GeV with longitudinal or transverse polarization [16]. Alwall, et al., proposed a basis of four simplified models for a first characterization of the new physics at the LHC, and they furnish a quantitative presentation of the jet structure, electroweak decays, and heavy-flavor content of the data, independent of detector effects [2]. Yang studied the radiative and leptonic decays of the pseudoscalar charmonium state $\eta_c$, which is sensitive to physics beyond SM, and measurement of the leptonic decay may give information about new physics [17]. Blum, et al., discussed the combination in $K^0 - \overline{K}^0$ and $D^0 - \overline{D}^0$ mixings to constrain the flavor structure of new physics [18]. IceCube Collaboration determined the atmospheric neutrino flux and searched new physics with AMANDA-II [19]. Idilbi, et al., derived the factorization and resummation for single color-octet scalar production at the LHC, and applicable to any new physics model provided the dominant production mechanism is gluon-gluon fusion [20]. Osland, et al., discussed the spin and model identification of Z′ bosons at the proton-proton CERN LHC, where heavy resonances appearing in the clean Drell-Yan channel may be the first new physics to be observed [21]. Golowich, et al., discussed the amplitudes for both $D^0 - \overline{D}^0$ mixing and the rare decays $D^0 \to l^+ l^-$ in excess of SM prediction could identify new physics contributions [22].

For the theories beyond SM, Rajpoot discussed the neutral-current interactions beyond the standard $SU(2)_L \times U(1)$ model [23]. Haber and Kane searched supersymmetry as probing physics beyond SM [24]. Fishbane, et al., discussed the charge-vectorial chiral sets of fermions beyond SM [25]. Bigi and Cveti studied the forward-backward asymmetries for $e^+ e^- \to Z^0 \to b\overline{b}, t\overline{t}$ and physics beyond SM [26]. Cveti and Lynn studied some gauge structures beyond SM [27]. Anglin, et al., considered the polarized ep collisions as probes for CP violation from beyond SM [28]. Langacker, et al., provided the high-precision electroweak experiments as a global search for new physics deviated quantitatively SM [1]. Einhorn considered the screening of heavy scalars beyond SM [29].

Recently McKeen investigated light bosons, which can be found in large classes of theories beyond SM, with radiative Υ′(1) decays [30]. Gavela, et al., analyzed the large gauge invariant nonstandard neutrino interactions, and are shown to require at least two new mediator particles [31]. Yang analyzed the radiative and leptonic decays of the pseudoscalar charmonium state $\eta_c$, while measurement of the leptonic decay may give information about new physics beyond SM [32]. Using lattice simulations, Appelquist, et al., studied conformal behavior in SU(3) Yang-Mills theories, and discuss open questions and the potential relevance of the present work to physics beyond SM [33]. Hafermann, et al., presented a novel approach to long-range correlations beyond dynamical mean-field theory, and studied the efficient perturbation theory for quantum lattice models [34]. Hatanaka and Yang studied radiative and semileptonic B decays involving the tensor meson $K_2^*(1430)$ in SM and beyond [35]. Theories beyond SM include a number of new particles, some of which might be light and weakly coupled to ordinary matter, Krivoruchenko, et al., discussed constraints for weakly interacting light bosons from existence of massive neutron stars [36]. Roy, et al., shown that how string resonances, whose formation is an exciting possibility of discover new physics beyond SM, may be detected at the LHC in the $pp \to \gamma + $ jet channel [37]. Jung, et al., performed the comparison of electric dipole moments and LHC for probing CP violation from physics beyond SM in triple boson vertices of the electroweak theory [38].

Faced to these complex theories, firstly we must distinguish carefully the primary theories, which are proved by some basic experiments, and other secondary theories, which are only mathematical or physical hypothesis or deduction. Next, those secondary theories should be discussed again. Now we propose ten bigger questions:

**1. Quark**

Hadrons possess very good symmetry from SU(3) to $SU_c(3) \times SU(6)$, etc., in particular, at low energy. At the present, these symmetries and corresponding characteristics are belonged to various quarks. Now various experiments have proved that smaller partons, which are asymptotically free, exist inside proton and all hadrons. Partons are quarks, which is namely quark-parton model (QPM). However, so far any quark is not found. What prove the existence of quark for the permanent quark confinement? While some unexpectedly properties are observed. Such partons may not be quarks.

Many experiments show that the masses of u and d quarks are small, the asymptotically free represents that interactions among quarks may be neglected. Quark model is the best simple, but may not be true. A puzzle is that the dynamical span is about 100000 times from u to t quarks. Their quarks are the same for both baryons $\Lambda = uds = \Sigma^0$, and only isospins I are different. But, both are very different. This is also inconsistent with a simple quark model.

When the same quarks form the hadrons, a binding energy ($m_u + m_d - m_\pi$) of $\pi$ meson composed of two quarks is very bigger than the binding energy ($2m_u + m_d - m_p$) of proton composed of three quarks. But, a life of meson is very short than proton. Assume that the mass of the u and d quarks is about 320MeV, the binding energy of meson 500MeV will be 25 times more than one of nucleon 20MeV. Now assume that the quark mass m(u)=1.5~3.0MeV, and m(d)=3~7MeV [39]. So a proton will be able not to be composed of three quark, while be composed of about 300 quarks.

Quarks and their composed ways may be different. For example, quarks may have three generations, and add different quark-pair $q\bar{q}$. They are $c = ud\vec{d}$, $s = du\vec{u}$, or $dd\vec{d}$, or mix-state $(d\vec{d} + u\vec{u})$, $t = uu\vec{u}$ or $ud\vec{d}d\vec{d}$ and so on. If $K^- = \bar{u}s = \bar{u}d(u\bar{u}) = \pi^-\pi^0$, later either mesons will be able to belong to SU(2). In theory the number of quarks in a hadron is invariant. But, in experiments the number of partons in a hadron is variable.

Of course, so long as quarks have invariant charges and changeable masses, the quark model is namely a type of structures, or quarks are the mathematical signs or partons, and may unify various hadrons.

These are inconsistent with the traditional common scientific knowledge, and support my hypothesis, in which quarks are possibly magic, and a substeady bound state with some symmetries [40]. We proposed correspondingly a possible model: Assume that all particles consist of numerous sandons. The Buddhist scripture says, as numerous as the sands of the Ganges. When the phase transition appears, prepreons, subquarks, quarks and particles at different shell are formed from sandons. Along with increase in energy, the various shell are opened incessantly, a particle can exhibit in many states. This is the so-called many shell-state model. Further, we combined the statistical sandons at high energy with the symmetrical magic quarks at low energy, and suggested a new symmetry-statistics duality of particle. The present theories and models, etc., of particle seem to be divided into symmetrical and statistical. It is possible development of the wave-particle duality, and possesses wide meaning [40].

We may also discuss a sandon-soft matter model, in which much sandons form soft-matter of particles, which corresponds to the fluid model. The flowing grain may produce the like-soliton. This corresponds to the quark-parton, i.e., quarks are formed when sandons flow or move.

The asymptotically free is proved by experiments [41,42], but the precondition of the quark confinement is the existence of quarks. The gluons are introduced in order to describe interaction among quarks, which should be a short-range interactions, and mass of gluon should be very big. But, the interaction is asymptotically free, and a gluon is color and massless. It is not obey with usual Yukawa interactions. If quarks do not exist, gluons will be corrected.

In lattice theory quark cannot be confined, so it should be discovered, and pass through phase transition to decompose. In the string model quark is similar with monopole. But, the usual magnetic field originates from the movement of charge and change of the electric field. Such quarks originate from the symmetry, or are also some certain movement states. Quark and monopole all are magic, both are similar each other, but do not exist in fact.

Einstein proposed an idea: a wave-field of photon may regard as a phantom field. This is similar that quark is only a type of magic particle, and is possibly a substeady state, and a probabilistic particle under a certainty energy-scale. Quarks show symmetry, and the distribution density of matter inside particle. It is an extension and generalization of Born probabilistic interpretation on wave function and of Einstein field on photons. Quarks only represent our knowledge concerning the particle-matter at present.

Quark is probably a phenomenological representation of the basic dynamics, and is a magic particle. If quarks are steady, they should exist. If quarks are instable, they will be not the true elements, whose quantum number should be conservation.

A name "quark" origins from James Joyce's novel *Finnegans Wake* [43]. It is a cry of gull, which cannot be seized, but may listen, and sound wave is also a type of moving state. Initially, Gell-Mann think, quark does not exist in fact. This seems to be a prophesy full wisdom: quark is not a true substance. Such a question of quark confined will be solved radically.

**2. Asymmetries of quantum numbers in the Standard Model**

The success of the Standard Model (SM) as a description of the properties of particles is well-known, so SM is believable [44]. But, it possesses some questions, for example, SM contains 19 parameters which have to be taken from experiment. Even to determine these parameters from experiment, except for the lepton masses, is a difficult and theory-laden task [45]. Moreover, the masses of hadrons cannot be obtained exactly from 19 parameters, and the quark mass has a region, for example, m(u)=1.5~3.0MeV, and m(d)=3~7MeV [39]. Therefore, SM is not a fundamental theory.

At present SM is a symmetry of the three generations of quark-lepton:

$$\begin{pmatrix} u \\ d \end{pmatrix} \sim \begin{pmatrix} \nu_e \\ e^+ \end{pmatrix}; \begin{pmatrix} c \\ s \end{pmatrix} \sim \begin{pmatrix} \nu_\mu \\ \mu^+ \end{pmatrix}; \begin{pmatrix} t \\ b \end{pmatrix} \sim \begin{pmatrix} \nu_\tau \\ \tau^+ \end{pmatrix}. \qquad (1)$$

Their charges are

$$\begin{pmatrix} 2/3 \\ -1/3 \end{pmatrix} \sim \begin{pmatrix} 0 \\ 1 \end{pmatrix}, Q(u) + Q(\nu_e) = Q(d) + Q(e^+) = \frac{2}{3}. \qquad (2)$$

Moreover, Neubert reviewed heavy quark symmetry [46]. But, this beautiful model implied some contradictions on symmetries. It deviates from the SU(3) symmetry, in which quarks are (u,d,s), so corresponding leptons should be ($\nu_e, e^+, \mu^+$), but there is not $\nu_\mu$. If the symmetrical group is SU(4), the three generations will be SU(6). For the symmetry on quark-lepton it forms only six symmetrical SU(2) groups. According to the complete symmetry, on the analogy of an isospin doublet u-d (I=1/2), we obtain various symmetrical isospin doublets: c-s, t-b, and three generations of lepton ($\nu_e - e, \nu_\mu - \mu, \nu_\tau - \tau$). But it is different that s and c quarks are two isospin singles (I=0), and there should be $I(\mu) = I(\nu_\mu) = 0$. In SM there are some basic differences with the former conclusions [47]. It is also a symmetric breaking.

### 3. Point particle, strength and distance of interactions

Quantum mechanics and quantum field theory are all based on the pointlike structure, and corresponding models and local interactions. At present various experiments proved that smaller partons exist inside proton and all hadrons. Such quantum theory must be developed.

Now a developed direction of experiments is mainly higher energy, so it should correspond to smaller space and particles. But, particles at higher energy, in particular, mesons, are probably very much, since mesons correspond to various strong interactions, and masses of particles and mesons are bigger at higher energy, and number increases. The collisions of particles at higher energy are analogous to cluster collision. It can obtain a composite state of middle process, and produce only new more and more heavy particles. But, this seems be not able to exhibit the internal structure of particles, as more heavy nuclei cannot redound to understand the internal structure of nucleon.

Further, quantum fluctuations at small distance will affect more and more interactions among particles, and bring ultra-violet divergence. In string theory the increase of energy is consumed almost to increase a scale of string, and higher energy detects bigger space. These experiments on partons explain that nucleons and hadrons collided by photons, neutrinos and electrons are more reason for the detection of hadronic structures. If the experiment on fractional charge of the nioblum ball is true, it will show powerfully that the fractional charge exists only at low energy, and quarks as component of hadrons should be small mass. These components agglomerate each other, and cannot be exhibited. Another direction should be to develop various detectors on weak information.

In particle physics these are some basic contradictions between short-range (strong and weak) interactions and interacting distances, and between coupled energy and stability on nucleon and meson, both consist of the same quarks [48].

Why the strong interaction cannot prevent particles decay under weak interaction? According to the descending relation $m(W,Z) > m(K) > m(\pi)$ of particle-masses on weak, medium strong and very strong interactions should assume that the strong interaction with attraction is very big at short distance, and the strong interaction is medium at shorter distance, and the strong interaction becomes weak interaction with repulsion at more shorter distance. But, the interaction among quarks with the shortest distance is again very strong. Singleton discussed a possibility that weak interaction becomes strong at 10 TeV [49].

**4. Mystic Higgs mechanism**

In SM Higgs bosons are responsible for generating the masses of all the elementary particles (leptons, quarks and gauge bosons) [50]. Their upper limit of mass calculated is

$$M_H < (4\sqrt{2}\pi/G)^{1/2} = 1.2 \text{TeV}. \tag{3}$$

Higgs mechanism is need for masses of hadrons. Further, it produces the gauge field and cosmic string, and obtains gravitation, and shows nuclear force and so on. Higgs bosons seem to be immanence, but they are the most untested one in SM [50].

Higgs bosons and two Higgs doublets exist in the minimal supersymmetric SM and the non-minimal supersymmetric model [51]. Recently Mao, et al., probed the coupling between the Higgs boson and vector gauge bosons and discovered the signature of new physics, and described the complete QCD corrections to associated Higgs boson production with a W boson pair at LHC [52]. Dawson and Yan considered models with multiple Higgs scalar gauge singlets, and emphasized the new physics resulting from the addition of at least two scalar Higgs singlets [53].

The based on the Lagrangian of symmetry and its dynamical breaking or Higgs breaking, we substituted the soliton-like solutions of the scalar field equations into the spinor field equations, and derive the Morse-type potential in the approximation of non-relativity, whose energy spectrum is the GMO mass formula and its modified accurate mass formula [40]:

$$M = M_0 + AS + B[I(I+1) - \frac{S^2}{2}]. \tag{4}$$

Further, according to the symmetry of s-c quarks, the heavy flavor hadrons which made of u,d and c quarks may be classified by SU(3) octet and decuplet, then we obtained correspondingly some simple mass formulas

$$M = M_0 + AC + B[I(I+1) - \frac{C^2}{4}], \tag{5}$$

or

$$M = M_0 + AC + B[I(I+1) - \frac{C^2}{2}]. \tag{6}$$

From this we predict some masses of unknown hadrons, for example, m($\Xi_{cc}$)=3715 or 3673MeV, and m($\Omega_{cc}^+$)=3950.8 or 3908.2MeV and so on [54]. Komamiya searched charged bosons at O(1/2-1 TeV) $e^+e^-$ colliders [50]. Higgs bosons may be something like the ether (the medium of light before the advent of special relativity), and not really exist [50].

## 5. Possible violation of the Pauli Exclusion Principle and some basic principles

The Pauli Exclusion Principle (PEP) is a basic one in quantum theory. In 1978, Santilli [55] was the first to propose the test of PEP, and then Ktorides, Myung and Santilli [56] pointed out the possible inapplicability of PEP under strong interactions, and possible deviations from PEP can at most be very small. In 1984, based on some experiments and theories of particles at high energy, I suggested that particles at high energy would possess a new statistics unifying Bose-Einstein (BE) and Fermi-Dirac (FD) statistics, and PEP would not hold at high energy [57,58]. In 1987, Ignatiev and Kuzmin constructed a model of a single oscillator violated PEP possibly. Then, Greenberg and Mohapatra [59,60] generalized to a local quantum field theory, where PEP should have a small violation. Such many very sensitive experiments are stimulated. But, so far these experiments show the validity of PEP with high precision for usual cases [61]. The future experiments should be combined widely with various theories of hidden and obvious violation of PEP, so some possible tests of the violation of PEP have been proposed [40,61-65].

If the condition of violation of PEP is high energy, it will corresponds to small space-time, and is related to the nonlinear theory [62], and the high density state, for example, plasma, quagma in nucleus, neutron stars and black hole, etc. If PEP is not hold under some conditions, quark model will keep simpleness, and need not color [58]. Such QCD will become a theory of strong interaction, in which these colorless quarks interchange gluons, and color charge is namely conservational quark with B=1/3.

Further, an extensive QCD as the universal strong interaction theory will be probably that the general baryons (fermions) pass through the vector fields to interact each other, in which the baryon-number B (fermion-number) is conservation, and corresponds to color. It possesses SU(N) symmetry.

Koures and Mahanthappra discussed the spontaneous parity violation in a supersymmetric nonlinear $\sigma$-model in 2+1 dimensions [66]. The violation of PEP contacts probably parity and CP violation and the PCT invariance. Ellis, et al., considered the violation of CPT and quantum mechanics in the $K_0 - \overline{K}_0$ system [67,68].

The possible violation of PEP corresponds to development of statistics and to the unified statistics [57]. On the other hand, possible decrease of entropy in an isolated system must have some internal interactions, in particular, for attractive process [69-71]. The basis of thermodynamics is the statistics, in which various interactions among the subsystems should not be considered [72]. The base of PEP is FD statistics, whose condition is also that interactions among particles are not considered. PEP has repulsion, and its violation should be attraction, for example, Cooper pair. If there are attractive forces in some systems, entropy will decrease, and PEP will be violated possibly, and FD and BE statistics will be unified. Both are related with black hole [58] and white hole. The violation of PEP and decrease of entropy should be consistent, and relevant to interactions and nonlinearity [62,71,73]. In a word, some basic principles are possibly correlated each other [65,74].

## 6. Neutrino and its oscillation

The neutrino physics is related to not only lepton physics, and limit velocity in relativity, and big bang cosmology and dark matter [75], and new physics [19], etc.

According to Yukawa interaction, massless neutrinos interchange W-Z with huge masses. It is very surprised. So far fourth lepton and a decay $\tau \to \mu\gamma$ are not discovered. Neutrinos should possess rest masses, and three kinds of neutrino $\nu_e, \nu_\mu, \nu_\tau$ should have three different rest masses. It is researched from some experiments and theories [76-80]. We proved that the Lorentz transformation is unsuitable for photon and neutrino, the photon transformation should be unified

$$x' = x + ct \text{ , (for space)} \qquad (7)$$

or $$t' = x'/c = t + (x/c) \text{. (for time)} \qquad (8)$$

It may reasonably overcome some existing difficulties, and the rest mass of photon and neutrino may be not zero [40,81].

In order to explain the solar neutrino problem, a pure theoretical hypothesis on neutrino oscillations is proposed. Such neutrinos must are with different rest masses (Dirac or Majorana mass), and they are the linear combinations of different masses for $|\nu_1>$ and $|\nu_2>$:

$$|\nu_e> = \cos\theta|\nu_1> + \sin\theta|\nu_2>, \qquad (9)$$

$$|\nu_\mu> = -\sin\theta|\nu_1> + \cos\theta|\nu_2>. \qquad (10)$$

Moreover, assume that the lepton-number is not conservation. It is relate to the supersymmetric grand unified theory with family symmetry [82], and large extra dimensions [83], etc. Gonzalez-Garcia and Nir reviewed the neutrino masses and mixing, and the phenomenology of neutrino oscillations in vacuum and in matter, and discussed the existence of new physics [84]. Recently Delepine, et al., computed the CP asymmetry in neutrino oscillations using the standard quantum field theory within a general new physic, and applied that the supersymmetry extensions of SM could produce a CP asymmetry observable in the next generation of neutrino experiments [85].

We think that the oscillations between neutrinos with huge different masses are very hardy. According to the symmetry of leptons, three neutrinos may not oscillate, and should be decay $\nu_\tau \to \nu_\mu \to \nu_e$, which is analogous to $\tau \to \mu \to e$, and even as $\mu$ and e cannot oscillate, only $\mu$ decays to e.

### 7. Uncertainty principle and superstring.

At present superstring is a very fashionable theory. A known size of superstring is very small $\Delta x \cong 10^{-35} cm$. According to the uncertainty principle, which is very important in quantum theory, a corresponding momentum of superstring is:

$$\Delta p = h/\Delta x \cong 6.626 \times 10^8 \, gcm/s \,. \qquad (11)$$

If the velocity of superstring is approximately velocity of light, its moving-mass will possess a macroscopic mass $5.618 \times 10^{26} MeV/c^2 = 2.209 \times 10^{-2} g$ [47]. Probably, superstring is only a graceful mathematical method. The scale of superstring exceeds greatly a scale of the great unification.

Recently the quantum entangled state is approved by some experiments. In these cases a difficulty is the uncertainty principle, which is probably breaking under some conditions.

Because there are weak interaction at $10^{-15}$ cm and corresponding repulsive force, perhaps hadrons cannot decompose to smaller particle as quark, the rather that superstring.

**8. The superposition principle and the entangled state**

The superposition principle (SP) lies at the foundation of quantum mechanics [86]. Various arguments between Einstein and Bohr, in particular, Einstein-Podolsky-Rosen argumentations (1935) are related to the linear superposition principle:

$$\Psi(x_1, x_2) = \sum \psi_n(x_2) u_n(x_1). \tag{12}$$

Many discussions, for example, Furry [87], Schrodinger [88] and Margenau [89] all are based on:

$$\Psi(x, y) = \sum c_n \gamma_n(x) \varphi_n(y). \tag{13}$$

Its development should combine the nonlinear quantum theory [40,65,90].

One of the basic hypothesis of von Neumann is:

$$<aR + bT + ...> = a<R> + b<T> + ... \tag{14}$$

and the operator quantity R+T+⋯ holds. They represent these linear relations, and derived the summability hypothesis, which is queried by G.Hermann (1935) and Jammer [91], since usual eigen value is not the linear combination.

Bohm, et al., proposed the possibility of the introduction of hidden variables in quantum mechanics based on the nonlinear relations, and derived a new deterministic equation of motion, describing a kind of coupling of the measuring instrument to the observed system that explains in detail how the wave packet is reduced during a measurement in a continuous and causally determined way [92]. But, Jauch and Piron opposed this correction [93]. Bell reconsidered the demonstrations of von Neumann and others, and proved that quantum mechanics is incompatible with the hidden variable interpretation [94].

Feynman had pointed out [95]: "In analysing physical phenomena we assume the validity of two principles: 1) special relativity, 2) the superposition of quantum mechanical amplitudes. As far as we know the principles are exact. The stage upon which all is played is the flat Minkowski space." Dirac [96] and Burt [97], et al., expounded the general SP. It holds always. But in the concrete cases the present applied SP is only a linear SP,

$$\psi = \sum_n C_n \psi_n. \tag{15}$$

Although various expressions of SP are different, but we require only: "It follows from the principle of superposition of states that all equations satisfied by wave functions must be linear in $\psi$" [98,99,96]. In fact many equations of wave functions already were nonlinear, for example, for some interactions. Various possible movement-states are namely various solutions of equations. The superposition of movement-states is namely the superposition of solutions. The plane wave may be superposition. But, the solitary wave and its equation cannot be linear superposition, so that particles corresponded to solitons cannot also be linear superposition. The wave and probability may together enter on a space. The solitary wave at a certain meaning is analogous to particle of wave and light. This shows that the superposition principle must be developed.

So long as "between these states there exist peculiar relationships", "state may be considered as the result of a superposition of two or more other states, and indeed in an infinite number of ways" [96]. This general SP includes the nonlinear superposition principle——Backlund transformation of soliton. Furthermore, in the nonlinear quantum theory the equations and operators are nonlinear [100], so the present applied linear superposition principle should be developed. If the linear SP connects with the probability interpretation of wave, this interpretation will be different for the quantum field theory. Further, we proposed a nonlinear quantum theory based on the nonlinear SP, and developed the linear operators and equations to nonlinear, and derived nonlinear Klein-Gordon equation and Dirac equations, and corresponding Heisenberg equation [40]. Many theories, models and phenomena are all nonlinear. Some known and possible tests for the nonlinear quantum theory are discussed [90].

Moreover, new experiments on the quantum nonlocality, the entangled state and teleportation [101-108] shown that the relation of quantum entangled state should be a new fifth interaction with middle strength and middle-rang [81,90]. Its characters are similar to the thought field [109]. The nonlinear SP is probably related to the correlations and the entangled state.

## 9. The wave property and duality

Based on the analysis of the logical structure of quantum mechanics, we prove that the wave-particle duality is the only basic principle of quantum mechanics. Statistics is the corresponding mathematical character [40]. This may be developed to the fermion-boson duality [110].

It is known that photons with higher frequency are more analogous to particles, so at high energy there is photon-photon interaction. Since the shape is invariant for collision of solitons, which cannot be the linear superposition, the nonlinear wave and solitary wave should be different with usual interference and diffraction. Soliton corresponds to particle, so the present wave property and the duality should be different. For single particle, or for short-range strong and weak interactions, which relate to few particles, or for small space-time, the probability-waves all are meaningless. For these cases the present wave property seems to be not able to be exhibited by present experiments [40]. Therefore, the essence of wave-property and the significance of wave function all must be new understanding [90]. Perhaps, the duality should develop to that wholeness are field, and a new symmetry-statistics duality [40].

## 10. Dark matter and dark energy

Astrophysical observations and cosmological arguments point out the necessity for a substantial amount of dark matter in the Universe. Many candidates of cold and hot dark matter are proposed, for instance, non-baryonic, weakly interacting massive particles (WIMPs) [111,112], the lightest supersymmetric particle [113,114], etc. Much theories on dark matter are researched [115-140].

Recently Belanger, et al., examined the predictions for both the spin-dependent and spin-independent direct detection rates in a variety of new particle physics models with dark matter candidates [141]. Cerde?o, et al., shown in a purely right-handed sneutrino can be a viable candidate for cold dark matter in the Universe in an extension of the minimal supersymmetric SM [142]. Sussman introduced quasilocal variables in spherical symmetry, and examined numerically the formation of a black hole in an expanding gas Friedman-Lema?tre- Robertson-Walker universe, as well as the evolution of density clumps and voids in an interactive mixture of cold dark matter and dark energy [143]. Gu, et al., suggested a hybrid seesaw model, and discussed neutrino masses, leptogenesis, and dark matter in the model [144]. Perez, et al., investigated a simple extension of SM, which predicts the existence of a pair of light charged scalars and, for vanishing triplet vacuum expectation value, contains a cold dark matter candidate [145]. Ettefaghi extended the noncommutative SM to incorporate singlet particles as cold dark matter [146]. Afshordi, et al., discussed the hierarchical phase space structure of dark matter haloes [147]. Bovy studied the substructure boosts to dark matter annihilation from Sommerfeld enhancement [148]. Henriques, et al., investigate a cosmological model based on the Salam-Sezgin six-dimensional supergravity theory, which include the unification of inflation, dark matter, and dark energy within a single framework [149]. McDonald, et al., presented an extension of the minimal supersymmetric SM include warm dark matter [150]. Sollom, et al., used cosmic microwave background radiation, large-scale structure and supernova data to constrain the possible contribution of cold dark matter isocurvature modes to the primordial perturbation spectrum [151].

We discussed again the Dirac negative energy state, which is a negative matter with the gravitation each other, and the repulsion with all positive matter. Such the positive and negative matters are two regions of topological separation in general case, and the negative matter is invisible. It is the simplest candidate of dark matter, and can explain some characteristics of the dark matter and dark energy. Based on a basic axiom and the two foundational principles of the negative matter, we research its predictions and possible tests. The negative matter should be a necessary development of Dirac theory. The existence of four matters on positive, opposite, and negative, negative- opposite particles will form the most perfect symmetrical world [152].

These ten examples show that particle physics are some remarkable questions on the best fundamental aspects. Here possible violation of PEP, the superposition principle and the wave property, etc., are contacted with the foundations of quantum mechanics. From new physics beyond SM to the entangled state and superluminal, and dark matter and dark energy, they indicate probably that a certain great fashion on physical foundation will arrive.